\documentclass[useAMS,usenatbib]{mn2e}
\usepackage{graphicx}
\usepackage{subfigure}

\def \OIII {[O{\sc iii}]5007~\AA}
\def \NII {[N{\sc ii}]6584~\AA}
\def \ha {H$\alpha$}
\def \SII {[S{\sc ii}]6717+6731~\AA}
\def \vhel{\ifmmode{~V_{{\rm HEL}}}\else{~$V_{{\rm HEL}}$}\fi}
\def\msun{\ifmmode{{\rm\ M}_\odot}\else{${\rm\ M}_\odot$}\fi}

\def\myr{\ifmmode{{\rm\ M}_\odot{\rm\ yr}^{-1}}
         \else{${\rm\ M}_\odot$ yr$^{-1}$}\fi}

\def\tena#1 #2 {\ifmmode{#1 \times 10^{#2}}\else{$#1 \times 10^{#2}$}\fi}
\def\kms{\ifmmode{~{\rm km\,s}^{-1}}\else{~km s$^{-1}$}\fi}

\title[Abell~41: shaping of a PN by a binary central star]{Abell~41: shaping of a planetary nebula by a binary central star?\thanks{Based on observations made with the William Herschel Telescope operated on the island of La Palma by the Isaac Newton Group in the Spanish Observatorio del Roque de los Muchachos of the Instituto de Astrof\'isica de Canarias.}}
\author[D. Jones et. al]{D. Jones,$^{1}$\thanks{E-mail:
david.jones-3@postgrad.manchester.ac.uk} M. Lloyd,$^{1}$ M. Santander-Garc\'ia,$^{2,3,4}$ J.A.L\'opez,$^{5}$ J. Meaburn,$^{1}$ 
\newauthor
D.L. Mitchell,$^{1}$ T.J. O'Brien,$^{1}$ D. Pollacco,$^{6}$ M.M. Rubio-D\'iez$^{7,2}$  and N.M.H. Vaytet $^{8}$
\\
$^{1}$Jodrell Bank Centre for Astrophysics, School of Physics and Astronomy, University of Manchester, M13 9PL, UK\\
$^{2}$Isaac Newton Group of Telescopes, Apartado de Correos 368, E-38700 Santa Cruz de La Palma, Spain\\
$^{3}$Instituto de Astrof\'isica de Canarias, E-38200 La Laguna, Tenerife, Spain\\
$^{4}$Departamento de Astrof\'isica, Universidad de La Laguna, E-38205 La Laguna, Tenerife, Spain\\
$^{5}$Instituto de Astronom\'ia, Universidad Nacional Aut\'onoma de M\'exico, Apartado Postal 877, 22800 Ensenada, B.C., M\'exico\\
$^{6}$Astrophysics Research Centre, Queen's University Belfast, BT7 1NN, UK\\
$^{7}$Departamento de Astrof\'isica, Centro de Astrobiolog\'ia, CSIC-INTA, Ctra. Torrej\'on a Ajalvir km.4, 28850 Madrid, Spain\\
$^{8}$Service d'Astrophysique, CEA/DSM/IRFU/SAp, Centre d'\'{E}tudes de Saclay, L'Orme des Merisiers, 91191 Gif-sur-Yvette Cedex, France}

\begin{document}

\date{Accepted xxxx xxxxxxxx xx. Received xxxx xxxxxxxx xx; in original form xxxx xxxxxxxx xx}

\pagerange{\pageref{firstpage}--\pageref{lastpage}} \pubyear{2010}

\maketitle

\label{firstpage}

\begin{abstract}
We present the first detailed spatio-kinematical analysis and modelling of the planetary nebula Abell~41, which is known to contain the well-studied close-binary system MT Ser.  This object represents an important test case in the study of the evolution of planetary nebulae with binary central stars as current evolutionary theories predict that the binary plane should be aligned perpendicular to the symmetry axis of the nebula.

Deep narrowband imaging in the light of \NII{}, \OIII\ and \SII{}, obtained using ACAM on the William Herschel Telescope, has been used to investigate the ionisation structure of Abell~41.  Longslit observations of the \ha\ and \NII\ emission were obtained using the Manchester Echelle Spectrometer on the 2.1-m San Pedro M\'artir Telescope.   These spectra, combined with the narrowband imagery, were used to develop a spatio-kinematical model of \NII\ emission from Abell~41.  The best fitting model reveals Abell~41 to have a waisted, bipolar structure with an expansion velocity of $\sim$40\kms{} at the waist. The symmetry axis of the model nebula is within 5$\degr$ of perpendicular to the orbital plane of the central binary system.  This provides strong evidence that the close-binary system, MT Ser, has directly affected the shaping of its nebula, Abell~41.

Although the theoretical link between bipolar planetary nebulae and binary central stars is long established, \emph{this nebula is only the second to have this link, between nebular symmetry axis and binary plane, proved observationally}.

\end{abstract}


\section{Introduction}

\begin{figure*}
\centering
\includegraphics{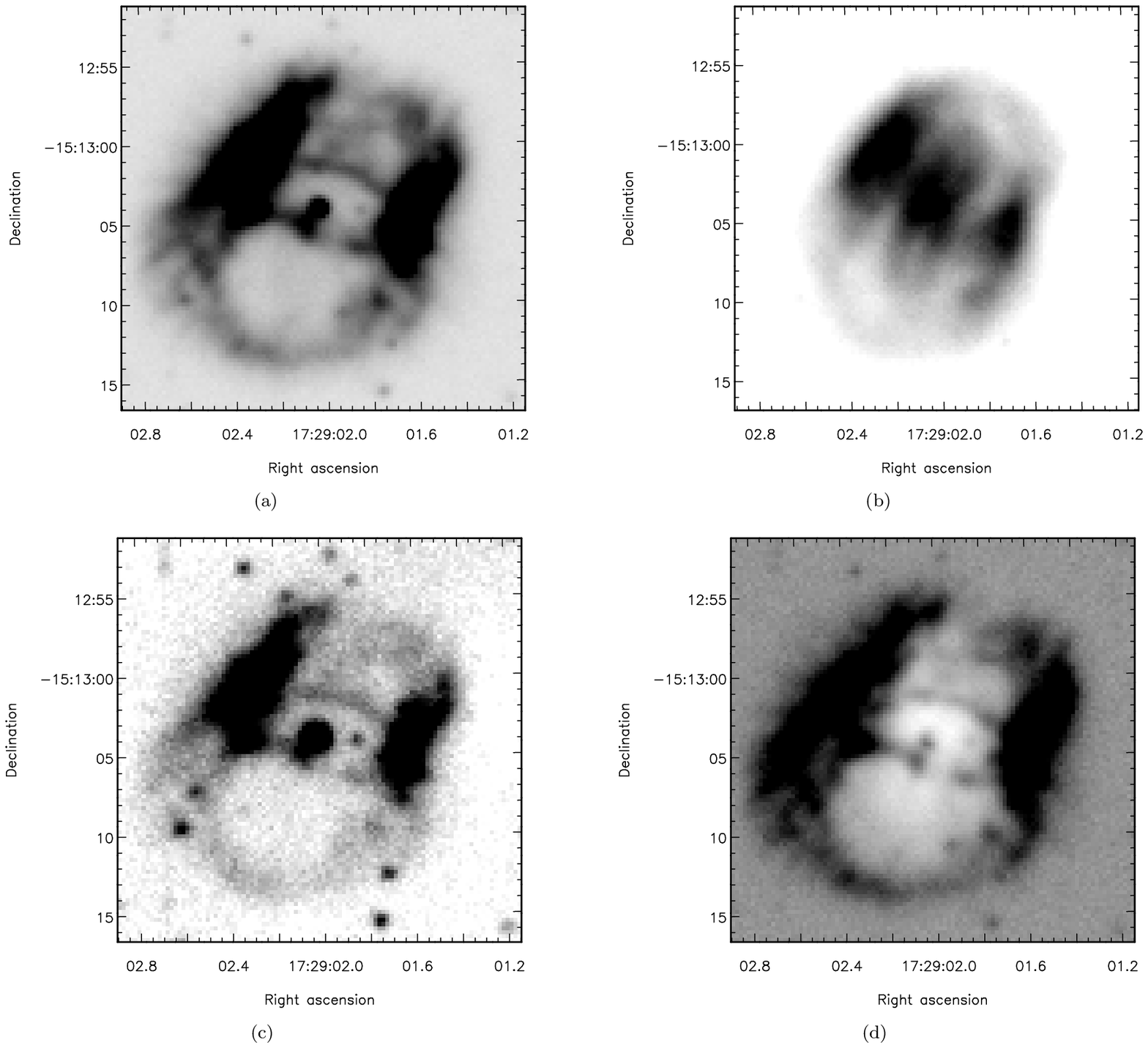}
\caption{Deep ACAM-WHT images of A~41 in the light of (a) \NII{}, (b) \OIII{}, (c) \SII\ and (d) \NII\ divided by \OIII{}, where black indicates a large ratio. Display scales have been chosen so as to highlight structural features referred to in the text.}
\label{fig:images}
\end{figure*}

It has long been believed that planetary nebulae (PNe) result from an early dense stellar wind, originating from the progenitor asymptotic giant branch (AGB) star, being swept into a thin shell by a later fast wind from the emerging white dwarf (WD).  \citet{kahn85} later extended this model, first developed by \citet{kwok78}, showing that an equatorially-enhanced AGB wind would lead to the formation of an aspherical nebula.  This model went on to become known as the generalised interacting stellar winds (GISW) model (see e.g., review by \citealp{balick02}), though GISW is not always consistent with observation of expansion velocity versus ionization structure (see e.g., \citealp{meaburn05b}).  The mechanism causing this anisotropic mass-loss is not yet clear.  The most widely accepted source of this anisotropy is for the central star of the planetary nebula (CSPN) to interact with a binary partner \citep{demarco09}.  To produce the most extreme bipolar PN structures (e.g. that of MyCn~18, the Etched Hourglass Nebula, \citealp{bryce97}), it is believed that the CSPN would have to form a common envelope (CE) with its binary companion \citep{bond90}.  For a CE to form, one component of the binary must accrete mass from its Roche lobe-filling partner more rapidly than it can thermally adjust to the additional material, thus also filling its Roche lobe.  After this point, any further mass lost by the donor will go on to form a CE surrounding both stars, and it is this CE which goes on to form the high equatorial density required by the GISW model \citep{nordhaus06}.  This mechanism is seen as the most likely method by which very close binaries (for example, cataclysmic variables) are formed \citep{grauer83}, because as the common envelope is ejected by transfer of angular momentum, the binary spirals in, dramatically shortening the period of the system.

Abell~41 (PN~G009.6+10.5, \mbox{$\alpha = 17^h29^m02.03^s$}, \mbox{$\delta = -15\degr13'04.4''$} J2000), discovered by \cite{abell66}, was classified by \citet{bond90} as elliptical under the classification scheme of \citet{balick87}.  However, deeper \ha{}$+$\NII\ imagery reveals ``that the nebular morphology exhibits an `H' shape with the addition of fainter material forming a continuous loop'' \citep{pollacco97}.

Photometric analysis of the CSPN, MT Ser, revealed it to be a close binary, showing minima at regular intervals of $2^h43^m$ \citep{grauer83}.  \citet{bruch01} confirmed the binary nature of MT Ser but were unable to accurately determine its orbital parameters because they found two different models which fit the observed data. (a) The binary consists of a hot sub-dwarf and a less evolved secondary, in which case the period is $2^h43^m$ and the variations are due to a reflection effect (inclination, $i = 42.52\degr \pm 1.73 \degr$) \footnote{Here, the inclination, $i$, is defined such that for $i=90\degr$ the orbital plane would be in the line of sight (i.e. eclipsing).}.  (b) The binary consists of two evolved, hot sub-dwarfs with a period of $5^h26^m$ where the variability results from partial eclipses and ellipsoidal variations ($i = 65.7\degr \pm 0.9\degr$).  They determined the optimum parameters for each model, but concluded that only radial velocity observations would be able to distinguish between the two.  Subsequent observation and modelling by \citet{shimanskii08} confirmed the presence of two sub-dwarf components, but gave no independent confirmation of the orbital inclination.    \citet{ogloza02} presented photometric and radial velocity observations along with modelling of MT Ser, independently determining an orbital inclination of $72\degr \pm 15\degr$, which is consistent with the second model of \citeauthor{bruch01} (\citeyear{bruch01}, $i = 65.7\degr \pm 0.9\degr$).  However, the values determined for the rest of the system parameters differ vastly, indicating that any agreement should be treated with caution.  Additionally, the stellar temperatures derived by \citet{ogloza02} are inconsistent with the observational detection of two hot sub-dwarf binary components \citep{shimanskii08}.  Of the two models of \citet{bruch01} and the model of \citet{ogloza02}, only the second model of \citeauthor{bruch01} (\citeyear{bruch01}, $i = 65.7\degr \pm 0.9\degr$) is consistent with photometric observations and the detection of two hot sub-dwarf central stars, indicating that this is the most reliable modelling of the binary CSPN system.

In this paper we present longslit spectroscopy of A~41, from which, combined with narrowband images, we derive a spatio-kinematical model of the nebula, with the aim of understanding the relationship between the nebula and MT Ser.

\begin{figure}
\centering
\includegraphics{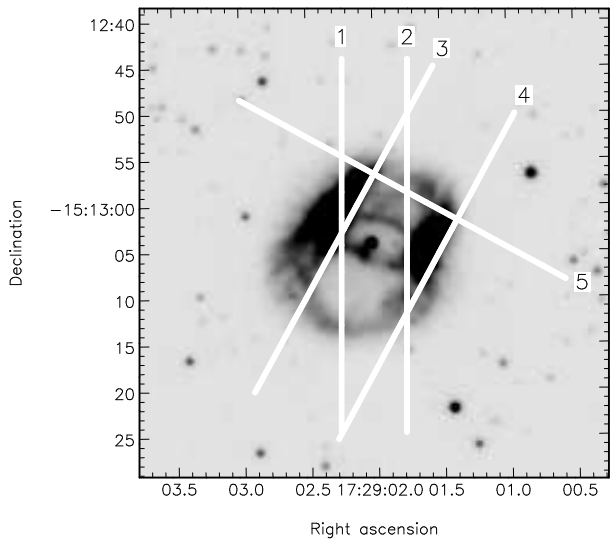}
\caption{A deep ACAM-WHT image of A~41 in the light of \NII\ showing the positions of the 5 MES-SPM longslits.}
\label{fig:slitimage}
\end{figure}

\begin{figure*}
\centering
\includegraphics{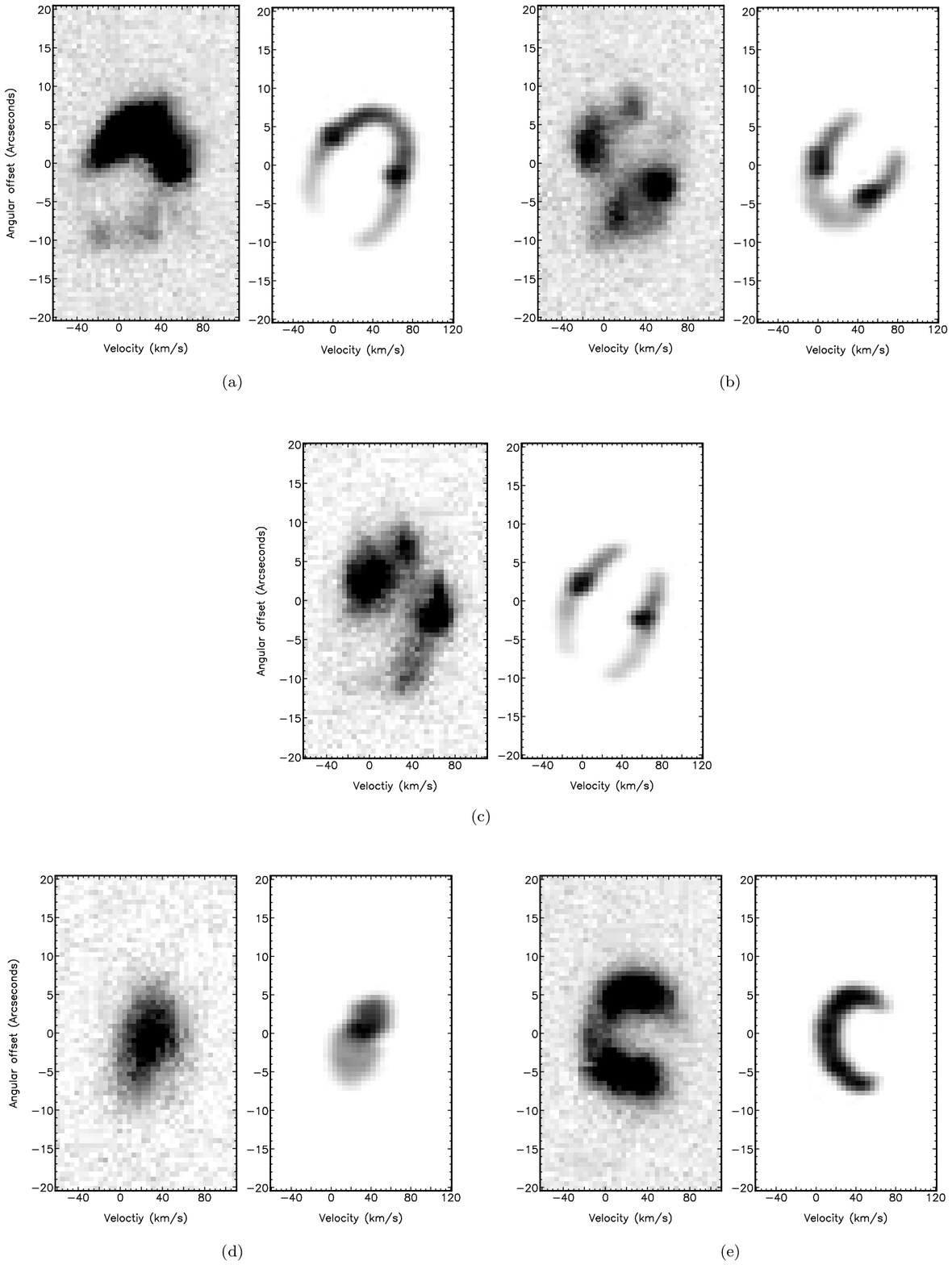}
\caption{Comparison of observed \NII\ PV arrays and synthetic equivalents: (a) observed and synthetic PV arrays of Slit 1, (b) observed and synthetic PV arrays of Slit 2, (c) observed and synthetic PV arrays of Slit 3, (d) observed and synthetic PV arrays of Slit 4 and (e) observed and synthetic PV arrays of Slit 5.  The slit positions are as shown in Figure \protect{\ref{fig:slitimage}}, and the velocity axis on all plots is heliocentric velocity, $V_{hel}$.}
\label{fig:pvarrays}
\end{figure*}

\section{Observations and Data Reduction}
\label{sec:obs}

Narrowband images of A~41 were acquired using ACAM combined with the 4.2-m William Herschel Telescope on 2009 August 4 (\NII\ and \OIII{}) and 2009 August 29 (\SII{}).  The seeing for both sets of observations did not exceed 0.9\arcsec{}.  ACAM was employed in standard imaging mode without binning resulting in a pixel scale of \mbox{$0.25 \arcsec$ pixel$^{-1}$}, and using the [NII]6584/21, Taurus 5009/15 and [SII]6727/48 filters (ING filters \#85, \#108 and \#86).  Three 15 minute exposures were taken in each filter, the data were bias-corrected, flat-fielded and cleaned of cosmic rays using \textsc{starlink} software.  \textsc{starlink} software was also used to remove the background lunar contamination, arising due to the close proximity of A~41 to the Moon, from the \SII\ images.  The resulting images were then co-added and are shown in Figure \ref{fig:images}.

Spatially resolved, longslit emission line spectra of A~41 were obtained with the second Manchester Echelle Spectrometer combined with the 2.1-m San Pedro M\'artir Telescope (MES-SPM, \citealt{meaburn03}).  MES-SPM was used in its primary spectral mode with a narrow-band \mbox{90 \AA{}} filter to isolate the \ha\ and [N~{\sc ii}] 6548 and 6584~\AA\ emission lines of the 87th echelle order.  Observations took place on two separate runs in 2004 June and 2007 June, both with the same instrument and set-up, using a SITe3 CCD with 1024 $\times$ 1024 24 $\mu$m square pixels ($\equiv 0.31\arcsec$ pixel$^{-1}$).  All integrations were of 1800 seconds.

Binning of 2 $\times$ 2 was adopted for all the spectral observations, resulting in 512 pixels in the spatial direction ($\equiv 0.62 \arcsec$ pixel$^{-1}$) and 512 pixels in the spectral direction ($\equiv 4.79 \kms$ pixel$^{-1}$).  The slit used was 30-mm long ($\equiv 5 \arcmin$) and 150 $\mu$m wide ($\equiv 2.0 \arcsec$ and $15\kms$).

Data reduction was performed using {\sc starlink} software.  The spectra were bias-corrected and cleaned of cosmic rays.  The spectra were then wavelength calibrated against a ThAr emission-lamp.  Finally, the data were rescaled to a linear velocity scale (relative to the rest wavelength of \NII{}, taken to be 6583.45 \AA{}) and corrected to show heliocentric velocity, $V_{hel}$.

In total, five integrations were obtained, two with slits running North-South, one with a position angle (PA) of 61.8\degr and a further two with PA equal to 151.8\degr.  The slit positions are shown in Figure \ref{fig:slitimage}, and the spectra are shown in Figure \ref{fig:pvarrays}.

\section{Analysis}
\subsection{Ionisation structure}
\label{sec:ionisationstructure}

Figure \ref{fig:images} shows A~41 in the light of \NII\ (a), \OIII\ (b) and \SII\ (c).  In all three emission lines the nebula displays roughly the structure remarked upon by \cite{pollacco97}, an ellipse where the western and eastern edges (parallel to the major axis of the apparent ellipse) appear much brighter than the rest of the nebula.  The northern side of the nebula also appears brighter than the southern side at all three wavelengths.  There are, however, differences between its appearance in the three emission lines; for example the ring-like feature at the centre of the nebula (the horizontal in the H analogy from \cite{pollacco97}), is much more well defined in \NII\ and \SII\ than in \OIII{}.  From the imaging alone it is unclear whether this ring is actually a material ring or a projection effect, where the ring is actually where the two lobes overlap in the line-of-sight.  It is also unclear whether the lobes themselves are open-ended or closed, as although their edges appear brighter (indicative of a closed shell, due to more nebular material in the line-of-sight), this effect is also seen in open-ended nebulae (such as MyCn~18: \citealp{sahai99a})

The nebula appears comparatively brighter centrally in the light of \OIII\ compared to \NII\ and \SII\ as shown by the ratio image in in Figure \ref{fig:images}(d).  Upon further inspection, this central bright region in \OIII\ appears to be an internal bipolar structure, although it is not clear whether this is just a line-of-sight brightness effect or, in fact, a separate internal nebular shell such as those seen in other nebulae (e.g.  NGC 7009: \citealp{sabbadin04} and Hen 2-104: \citealp{santander08}).  We note that this feature can also be found in the H$\alpha$+[N{\sc ii}] image shown in \citeauthor{miszalski09b} (\citeyear{miszalski09b}, originally shown in \citealt{pollacco97}).

\begin{figure}
\centering
\includegraphics{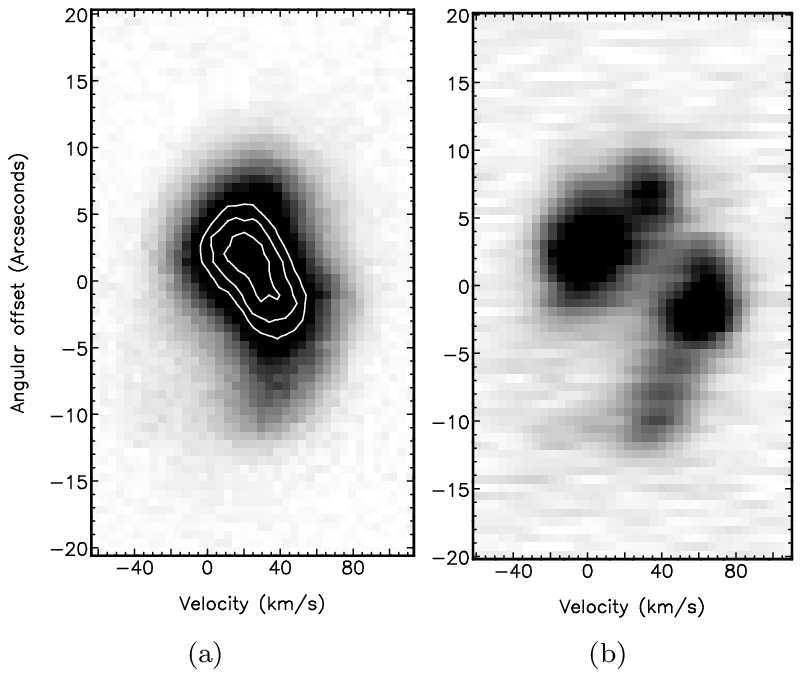}
\caption{PV arrays of Slit 3 (as shown in Figure \protect{\ref{fig:slitimage}}) in the light of (a) \ha\ with white contours marking the 95th, 97th and 99th percentiles and (b) \NII\ convolved to the same thermal width as that of \ha\ (assuming $T_e = 10^4$ K, the original \NII\ PV array is shown in Figure \protect{\ref{fig:pvarrays}}(c)).}
\label{fig:thermal}
\end{figure}

The nebular position-velocity (PV) profiles also appear different in the light of \NII\ compared to \ha{}, even once the different thermal widths of the two lines have been taken into account (see Figure \ref{fig:thermal}).  This ionisation stratification is a common feature among planetary nebulae, where the \NII\ emission is seen to trace the outside of the nebular shell whereas the \ha\ is relatively evenly distributed throughout (see e.g., \citealp{gurzadyan97}).

The nebular PV arrays also show the same brightness variations across the nebula as the imaging, for example Slit 1 [Figure \ref{fig:pvarrays}(a)] shows a bright partial velocity ellipse from the northern part of the nebula but much fainter emission from the southern part.  Slit 3 [Figure \ref{fig:pvarrays}(c)], which lies approximately parallel to the major axis of the apparent nebular ellipse, shows that the northern end of the nebula is red-shifted with respect to the south indicating that the nebular inclination is such that the northern side of the nebula is pointed away from the observer and the south towards.  The Slit 3 echellogram also shows two separate emission regions from the near and far sides of the nebular shell (i.e. no closed velocity ellipse), suggesting that the nebula is open at both ends of this symmetry axis. This is supported by the emission profile from, the slit perpendicular to Slit 3, Slit 5 [Figure \ref{fig:pvarrays}(e)], which shows a partial velocity ellipse.  The ellipse is not closed as a result of both the open ended and inclined nature of the nebular shell, meaning that the slit cuts across a region where only the blue-shifted side of the nebula is present. However, projection effects, surface brightness and shell thickness variations could produce an apparently open velocity ellipse from a closed shell, due to the sensitivity limits of the observations.

The intrinsic nebular structure and variation in brightness across the nebula are discussed further in Sections \ref{sec:modelling} and \ref{sec:ism}.

\subsection{Spatio-kinematical modelling of Abell 41}
\label{sec:modelling}

\begin{figure}
\centering
\includegraphics{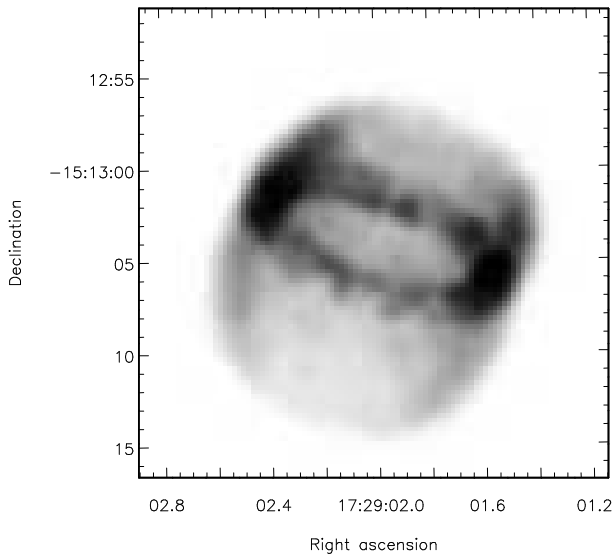}
\caption{The synthetic \textsc{shape} model for the NII emission from A~41 displayed at the same scale as those images in Figure \protect{\ref{fig:images}}.}
\label{fig:modelimage}
\end{figure}

\begin{figure}
\centering
\includegraphics{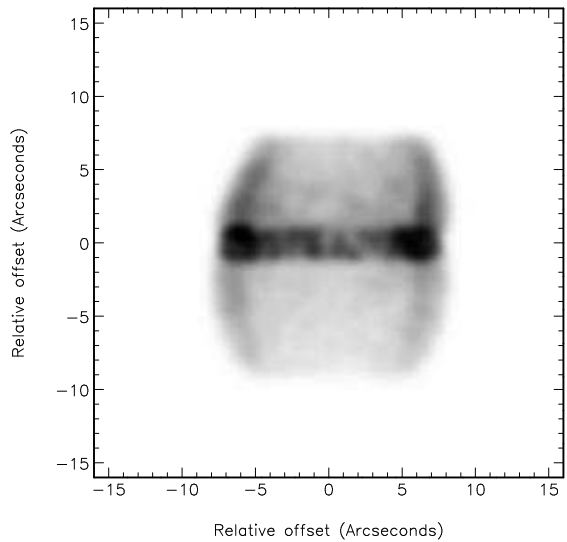}
\caption{The synthetic \textsc{shape} model for the NII emission from A~41 displayed at an inclination of 90\degr.}
\label{fig:modelsideon}
\end{figure}

A spatio-kinematical model, corresponding to the simplest three-dimensional structure consistent with the large-scale nebular \NII\ emission features, has been derived for A~41.  \NII\ emission was selected rather than \ha\ emission due to its lower thermal broadening and its shell-like distribution (as discussed in section \ref{sec:ionisationstructure}).  The modelling was performed in order to confirm the bipolar nature of the nebular shell and to constrain the inclination angle of this shell, for comparison with the inclination of the central binary (MT Ser).  The model was developed using \textsc{shape} \citep{steffen06} and by assuming a Hubble-type flow, where expansion velocity is radial and proportional to the distance from the centre of the nebula.  The model parameters (dimensions, shape, expansion scale velocity and inclination) were manually varied over a wide range of values and the results compared by eye to both spectral observations and imaging, until a best-fit was found.  This best fit model comprises a bipolar shell waisted by an equatorial ring with an expansion velocity of $\sim$40\kms{}.  The model nebula is slightly asymmetric in that the northern lobe is shortened by 15\%, has a narrower opening angle and has a slight shear with respect to its southern counterpart. No symmetric model could be found to reproduce the observed PV arrays. The nebular inclination angle, as defined by the un-sheared southern lobe, is determined to be $66\degr \pm 5\degr$ (in excellent agreement with the value determined by \citeauthor{pollacco97}, \citeyear{pollacco97}, by deprojecting the nebular ring). The model nebula is shown at the observed orientation in Figure \ref{fig:modelimage} and at an inclination of 90\degr{}, to highlight the asymmetry, in Figure \ref{fig:modelsideon}.  The synthetic PV arrays are shown, along with their observed counterparts, in Figure \ref{fig:pvarrays}.

It is not unheard of for bipolar nebulae to show asymmetry between the opposing lobes, both in the extension of the lobes (e.g. NGC 6881: \citealp{ramos-larios08}) and their opening angles (e.g. OH231.8+4.2: \citealp{sanchezcontreras04}).  In comparison, the level of asymmetry shown by A~41 is very low.  One possible explanation for the asymmetry is discussed in Section \ref{sec:ism}.

Also of note, the equatorial expansion velocity, of $\sim$40\kms{}, is unusually large for a bipolar nebula, though not exceptionally so (\citealt{solf84}, \citealt{weinberger89}).

\subsection{Systemic velocity and kinematical age}

Comparison of synthetic model spectra to their observed counterparts provides an unambiguous measure of the nebular systemic heliocentric radial velocity ($V_{sys}$), unaffected, for example, by brightness variations or nebular asymmetry \citep{jones10}.  Using the best-fit model described in Section \ref{sec:modelling}, $V_{sys}$ is determined to be 30 $\pm$ 5 \kms\ in good agreement with the value of 30 \kms\ determined by \citet{beaulieu99}.  Similarly, the nebular expansion velocity, determined by the kinematical modelling, can be used to calculate a kinematical age for the nebula.  This, however, requires the distance to the nebula to be known.  The distance to A~41 is a matter of some debate with values in the literature ranging from $\sim$1 kpc \citep{grauer83} up to 9.0 $\pm$ 0.4 kpc \citep{shimanskii08}, this probably results from the notorious variation in results from different methods of distance determination (see e.g., \citealp{gurzadyan97}).  Therefore, rather than favour one particular distance estimate over another we quote a kinematical age per kiloparsec of $\sim800$ years kpc$^{-1}$.

\subsection{Evidence of ISM interaction?}
\label{sec:ism}

The asymmetries in both brightness and shape (discussed in Sections \ref{sec:ionisationstructure} and \ref{sec:modelling}) could be thought of as strong evidence for interaction with the interstellar medium (ISM).  Consider, if the nebula were moving through the ISM with the northern lobe at the leading edge, it would be reasonable to expect to see this lobe brightened (through shock excitation) and less extended (due to the greater drag) with respect to its relatively unimpeded southern counterpart.  This brightening and lesser extension is precisely what is borne out by the data and subsequent modelling.  Similarly, if the motion of the nebula, with respect to the ISM, were not along the symmetry axis but slightly offset, then this would also offer an explanation for the ``shear" seen in the northern lobe (i.e. the shear is as a result of the eastern side of the lobe being impeded by the ISM more than the western side), and indeed the brightness contrast across that lobe (eastern edge brightened with respect to the western edge).  

We do not see any evidence of a bow-shock in any of our observations, which might be expected to be particularly prevalent in the \OIII\ image \citep{wareing07}.  However, the relatively low levels of asymmetry in the nebula indicate that any interaction is fairly weak, implying a slow relative velocity consistent with no bow shock.

The ratio of \SII\ to \NII\ provides a good tracer for shock excitation which might be expected to arise from interaction with the ISM \citep{phillips98}.  However, the close proximity of the Moon to A~41 during the \SII\ imaging detailed in Section \ref{sec:obs}, and the uncertain nature of the subtraction of this background, means we are unable to accurately assess the \SII\ to \NII\ ratio of this nebula. Similarly, given the proper motion and $V_{sys}$ of A~41 it would be possible to determine the motion of the nebula relative to the local ISM (assumed to be due to Galactic rotation, as in \citealp{meaburn09}).  The direction of this motion could then be used to assess the validity of this hypothesis, unfortunately we are unable to perform this analysis as no proper motion measurements exist for this nebula.  

Also, of interest is the apparent location of the central star, which one would expect to be offset towards the leading edge of the nebula by any interaction between nebula and ISM (in the case of A~41 offset roughly towards the North of the nebula). In fact, in our imagery it is slightly offset to the south of the centre of the nebular ring, at odds with this hypothesis.

\section{Conclusions}

Using high-resolution longslit spectroscopy and deep imaging, a spatio-kinematical model of A~41 has been developed which clearly shows that the nebula is aligned with the binary central system exactly as predicted by current theories of PN shaping by binary central stars.  This is only the second nebula to have this link observationally constrained (after A~63, \citealp{mitchell07b}).  The kinematical data confirm A~41 exhibits a waisted, bipolar structure, with some small deviations from perfect axisymmetry.  The presence of an equatorial ring is also confirmed, adding further weight to the link between ring structures and central binary stars as commented on by \citet{miszalski09b}.  

The data indicate that A~41 may be experiencing an interaction with the ISM and this too may be affecting its shape and brightness.

Further kinematical investigations, such as the one presented in this paper, coupled with in-depth studies of the CSPN of other PNe with confirmed close-binary central stars, are necessary to investigate the full extent of the influence of central star binarity on PN nebular shaping.  Only once a significant statistical sample has been acquired can generalisations be made about the role of CSPN binarity in PN evolution. 

\section*{Acknowledgments}

We would like to thank the anonymous referee for their comments, which greatly improved the clarity of this paper. We would also like to thank Andrew Cardwell and Pablo Rodr\'iguez-Gil for their assistance in the planning and acquisition of the WHT-ACAM imaging.  D.J. gratefully acknowledges the support of STFC through his studentship.  N.M.H.V. gratefully acknowledges financial support from ANR; SiNeRGHy grant number ANR-06-CIS6-009-01.

\bibliographystyle{mn2e}
\bibliography{literature.bib}

\label{lastpage}

\end{document}